\documentclass[12pt]{article}
\usepackage{amsmath,epsfig}
\input{epsf}
\def\ltap{\raisebox{-.4ex}{\rlap{$\,\sim\,$}} \raisebox{.4ex}{$\,<\,$}}

\newcommand\as{\alpha_{\mathrm{S}}}

\newcommand\f[2]{\frac{#1}{#2}}

\def\ms{${\overline {\rm MS}}$}
\def\beeq{\begin{eqnarray}}
\def\eeeq{\end{eqnarray}}

\def\to{\rightarrow}

\def\la{\lambda} 
\def\nn{\nonumber}

\def\chitwodof{\chi^2/{\rm d.o.f.}}

\begin{document}
\begin{titlepage}
\renewcommand{\thefootnote}{\fnsymbol{footnote}}
\begin{flushright}
CERN-PH-TH/2004-137\\
    hep-ph/0407241
     \end{flushright}
\par \vspace{10mm}
\begin{center}
{\Large \bf
The back-to-back region \\[0.8ex]
in $e^+e^-$ energy--energy correlation}

\end{center}
\par \vspace{2mm}
\begin{center}
{\bf Daniel de Florian$\footnote{Partially supported by Fundaci\'on Antorchas, UBACyT, ANPCyT and Conicet}
{}^{\,a}$}
\hskip .2cm
and
\hskip .2cm
{\bf Massimiliano Grazzini$
{}^{\,b}$}\\

\vspace{5mm}

${}^{a}\,$Departamento de F\'\i sica, FCEYN, Universidad de Buenos Aires,
(1428) Pabell\'on 1 Ciudad Universitaria, Capital Federal, Argentina

${}^{b}\,$Department of Physics, Theory Division, CERN, CH-1211 Geneva 23, Switzerland

\end{center}

\par \vspace{2mm}
\begin{center} {\large \bf Abstract} \end{center}
\begin{quote}
\pretolerance 10000

We consider the back-to-back region in the
energy--energy 
correlation in $e^+e^-$ collisions.
We present the explicit expression
of the ${\cal O}(\as^2)$ logarithmically enhanced contributions up to
next-to-next-to-leading logarithmic accuracy.
We study the impact of the results in a detailed comparison with precise
LEP and SLC data. We find that, when hadronization effects are
taken into account as is customarily done
in QCD analysis in $e^+e^-$ annihilations,
the extracted value of $\as(M_Z)$ is in good agreement with the current
world average.

\end{quote}


\vspace*{\fill}
\begin{flushleft}
     hep-ph/0407241 \\July 2004

\end{flushleft}
\end{titlepage}

\setcounter{footnote}{1}
\renewcommand{\thefootnote}{\fnsymbol{footnote}}

\section{Introduction}

Precise data on $e^+e^-$ annihilation into hadrons have provided detailed experimental tests of QCD and one of the best opportunities to measure the strong coupling constant $\as$. A particularly well suited observable is the energy--energy correlation (EEC) \cite{Basham}, defined as an energy-weighted correlation for the cross section corresponding to the process $e^+e^-\to h_a+h_b +X$ as
\begin{equation}
\label{eecdef}
\f{1}{\sigma_T}\f{d\Sigma}{d\cos\chi} = \f{1}{\sigma_T}
\sum_{a,b}\int \f{E_a\, E_b}{Q^2}d \sigma_{e^+e^-\to h_ah_b+X}\,
\delta(\cos\chi+\cos\theta_{ab})\, ,
\end{equation}
where $E_a$ and  $E_b$ are the energies of the particles, $Q$
is
the centre-of-mass energy,
$\theta_{ab}\equiv\pi-\chi$
is the angle between the two hadrons, and $\sigma_T$ is the total cross section
for $e^+e^-\to {\rm hadrons}$.

The two-hadron cross section 
$e^+  e^-\to h_a+h_b +X$  depends on the
fragmentation functions of the partons into the final-state hadrons.
However, thanks to the momentum sum rule
\begin{equation}
\sum_h\int_0^1 dx x D_{h/q}(x,\mu_F^2)=1\, ,
\end{equation}
EEC becomes independent of them, and can thus be computed
in QCD perturbation theory.

Theoretical calculations \cite{Stirling,Chao:1982wb,Schneider:1983iu}
for the EEC function have been performed up to next-to-leading order (NLO)
accuracy in QCD
\cite{Ali:1982ub}--\cite{lep2},
allowing a detailed comparison with the available data.

As is well known, fixed-order calculations have a limited kinematical
range of applicability. In the back-to-back region, defined by 
$\theta_{ab}\to \pi$ ($\chi\to 0$),
the multiple emission of soft
and collinear gluons
gives rise to large logarithmic contributions of the form
$\as^n \log^{2n-1}y$,
where $y=\sin^2\chi/2$.
As $y$ decreases,
the logarithms become large and therefore invalidate the use of
the fixed-order perturbative expansion.

These logarithmic contributions can be resummed
to all orders \cite{CollinsSoper,KodairaTrentadue}.
The resummation formalism is very close to the one developed
for the transverse-momentum distribution of high-mass
systems in hadronic collisions
\footnote{The role of the transverse momentum is played, in the case of EEC,
by the variable $q_T^2=Q^2\sin^2(\chi/2)$.}.
When the transverse momentum $q_T^2$
of the detected final state is much smaller
than its invariant mass $Q^2$, large logarithmic contributions 
$\as^n \log^{2n-1} q_T^2/Q^2$ arise which must be resummed to all orders.

The coefficients that control the resummation at a given order
can be computed if an analytical calculation at the same order exists.
In the case of hadronic collisions,
the complete form of the logarithmically enhanced contributions
has been computed \cite{b2calc}.
In this paper we present the result of a similar calculation, performed
for EEC. Our calculation allows us to fix the still missing coefficients at
${\cal O}(\as^2)$ and to extend the resummation for this observable to
full next-to-next-to-leading logarithmic (NNLL) accuracy.
We also study the numerical impact of our results and present a comparison
with LEP and SLC data.

The paper is organized as follows.
In Sect.~2 we review the resummation formalism and we discuss the results of
our calculation. In Sect.~3 we present
numerical results, and we also consider the inclusion
of hadronization effects.
In particular, we perform a fit to OPAL and SLD data.

\section{Soft-gluon resummation}

The EEC function can be decomposed as
\begin{equation}
\label{deco}
\f{1}{\sigma_T}\f{d \Sigma}{d\cos\chi}=
\f{1}{\sigma_T}\f{d \Sigma^{\rm (res.)}}{d\cos\chi}
+\f{1}{\sigma_T}\f{d \Sigma^{\rm (fin.)}}{d\cos\chi}\, ;
\end{equation}
the first term on the right-hand side of Eq.~(\ref{deco}) contains all the
logarithmically enhanced contributions, $\as^n/y \log^m y$ at small $y$ and has to be
evaluated by resumming them to all orders.
The second is free of such contributions and can be computed by using
fixed-order perturbation
theory.

The resummed component
can be written as \cite{CollinsSoper,KodairaTrentadue}
\begin{equation}
\label{CS}
\f{1}{\sigma_T}\f{d \Sigma^{\rm (res.)}}{d\cos\chi}=
\f{Q^2}{8} H(\as(Q^2))
\int_0^\infty\, db\, b\, J_0(bq_T)
S(Q,b)\,  .
\end{equation}
The large logarithmic corrections are exponentiated
in the
Sudakov form factor
\begin{equation}
\label{sudakov}
S(Q,b)=\exp \left\{ -\int_{b_0^2/b^2}^{Q^2} \frac{dq^2}{q^2} 
\left[ A(\as(q^2)) \;\ln \frac{Q^2}{q^2} + B(\as(q^2)) \right] \right\}\, .
\end{equation}
The Bessel function $J_0(bq_T)$ and $b_0=2 e^{-\gamma_E}$
have a kinematical origin.

The resummation formula in Eq.~(\ref{CS}) has a simple physical interpretation. 
When the triggered
partons are back to back, the emission of accompanying radiation
is strongly inhibited and only soft and collinear partons can be radiated.
The function $H(\as(Q^2))$ embodies hard contributions from virtual corrections at scale
$q\sim Q$. The form factor $S(Q,b)$ contains virtual and real contributions from
soft (the function $A$) and
flavour-inclusive
collinear (the function $B$) radiation at scales $1/b\ltap q \ltap Q$. At extremely low scales, $q\ltap 1/b$, real and virtual corrections cancel because EEC is infrared safe.

The functions $A$, $B$ and $H$ in Eqs.~(\ref{CS},\ref{sudakov}) are free of logarithmic corrections and can be computed using a perturbative
expansions in $\as$:
\begin{eqnarray}
\label{aexp}
A(\as) &=& \sum_{n=1}^\infty \left( \frac{\as}{\pi} \right)^n A^{(n)} \;\;, \\
\label{bexp}
B(\as) &= &\sum_{n=1}^\infty \left( \frac{\as}{\pi} \right)^n B^{(n)}
\;\;, \\
\label{cexp}
H(\as) &=& 1 + 
\sum_{n=1}^\infty \left( \frac{\as}{\pi} \right)^n H^{(n)} \;\;.
\end{eqnarray}

By explicitly performing the $q^2$ integration in Eq.~(\ref{sudakov})
the form factor can be recast in
the following form \cite{Turnock:wy,Frixione:1998dw,Catani:2000vq}:
\begin{align}
S(Q,b)=\exp\{L \, g_1(a_{\mathrm{S}} \beta_0 L ) + g_2(a_{\mathrm{S}} \beta_0 L ) + a_{\mathrm{S}}\, g_3(a_{\mathrm{S}} \beta_0 L )...\}\, ,
\end{align}
where $a_{\mathrm{S}}=\as/\pi$
and the large logarithm $L=\log Q^2 b^2/b_0^2$ at large $b$ corresponds to the
$\log y$, which becomes large at small $y$ (the limit $y\ll 1$ ($q_T\ll Q$) corresponds
to $Qb\gg 1$ through a Fourier transform).

The explicit expressions of the $g_i$ functions
are\footnote{Throughout the paper we use the \ms~renormalization scheme.}: 
\begin{align}
\label{gfunctions}
g_1(\la) &= \f{A^{(1)}}{\beta_0} \f{\la+\log(1-\la)}{\la} \nn \\
g_2(\la) &= \f{B^{(1)}}{\beta_0} \log(1-\la) -\f{A^{(2)}}{\beta_0^2} \left( \f{\la}{1-\la} +\log(1-\la)\right) + \f{A^{(1)}}{\beta_0} \left( \f{\la}{1-\la} +\log(1-\la)\right) \log\f{Q^2}{\mu_R^2}  \nn \\
& +\f{A^{(1)} \beta_1}{\beta_0^3} \left( \f{1}{2} \log^2(1-\la)+ \f{\log(1-\la)}{1-\la} + \f{\la}{1-\la}  \right) \nn \\
g_3(\la) &= -\f{A^{(3)}}{2 \beta_0^2} \f{\la^2}{(1-\la)^2}
-\f{B^{(2)}}{\beta_0} \f{\la}{1-\la}  
+\f{A^{(2)} \beta_1}{\beta_0^3} \left( \f{\la (3\la-2)}{2(1-\la)^2}
 - \f{(1-2\la) \log(1-\la)}{(1-\la)^2} \right) \nn \\
& + \f{B^{(1)} \beta_1}{\beta_0^2} \left( \f{\la}{1-\la} + \f{\log(1-\la)}{1-\la} \right) - \f{A^{(1)}}{2} \f{\la^2}{(1-\la)^2}  \log^2\f{Q^2}{\mu_R^2} \nn \\
&+ \log\f{Q^2}{\mu_R^2} \left( B^{(1)} \f{\la}{1-\la} + \f{A^{(2)}}{\beta_0}
 \f{\la^2}{(1-\la)^2} + A^{(1)} \f{\beta_1}{\beta_0^2}
 \left( \f{\la}{1-\la} + \f{1-2\la}{(1-\la)^2} \log(1-\la) \right) \right) \nn \\
&   +A^{(1)} \left( \f{\beta_1^2}{2 \beta_0^4} \f{1-2\la}{(1-\la)^2} \log^2(1-\la) 
+ \log(1-\la) \left[  \f{\beta_0 \beta_2 -\beta_1^2}{\beta_0^4} +\f{\beta_1^2}{\beta_0^4 (1-\la)}  \right] \right.  \nn \\
& \left. + \f{\la}{2 \beta_0^4 (1-\la)^2} ( \beta_0 \beta_2 (2-3\la)+\beta_1^2 \la)
 \right)        
\end{align}
and the coefficients of the QCD $\beta$ function are defined as:
\begin{align}
\beta_0 &= \frac{1}{12} \left( 11 C_A - 2 n_f \right) \;\;,
\quad\quad \beta_1=  \frac{1}{24} 
\left( 17 C_A^2 - 5 C_A n_f - 3 C_F n_f \right) \;\;,
\nonumber \\
\label{bcoef}
\beta_2 &= \frac{1}{64} \left( \f{2857}{54} C_A^3
- \f{1415}{54} C_A^2 n_f - \f{205}{18} C_A C_F n_f + C_F^2 n_f
+ \f{79}{54} C_A n_f^2 + \f{11}{9} C_F n_f^2 \right) \;\;.
\end{align}

The functions $g_1$, $g_2$, $g_3$ control the LL, NLL, NNLL contributions, respectively.
The coefficients $A^{(1)}$,  $A^{(2)}$ and $B^{(1)}$
were computed a long time ago \cite{KodairaTrentadue}
and are the same as appear in the quark form factor in the transverse
momentum distributions in hadronic collisions.
They read 
\begin{align}
\label{coef1}
A^{(1)}&= C_F \nn \\
B^{(1)}&= -\f{3}{2} C_F \nn \\
A^{(2)}&= 
\f{1}{2}\Big( C_A\,\left(\f{67}{18}-\f{{\pi }^2}{6}\right)-\f{5}{9}n_f \Big)\,\, A^{(1)} \, .
\end{align}
By using for $\sigma_T$ the NLO expression
\begin{equation}
\sigma_{T}^{\rm NLO}= \f{4\pi \alpha^2}{Q^2} \sum_q e_q^2  \left(1+\f{\as}{2\pi}\, \f{3}{2}\, C_F\right)\, ,
\end{equation}
the coefficient $H^{(1)}$ is \cite{Basham}
\begin{equation}
\label{H1}
H^{(1)}=  -C_F \left(\f{11}{4}+\f{\pi^2}{6}\right) \,.
\end{equation}
The coefficient $A^{(3)}$ has been obtained recently, as the leading soft term  in the three-loop splitting functions \cite{Moch:2004pa,Vogt:2004mw}:
\begin{align}
A^{(3)}=\f{1}{4}&\left[ C_A^2 \left( \f{245}{24} -\f{67}{9} \zeta_2 +\f{11}{6} \zeta_3 +\f{11}{5} \zeta_2^2 \right) +C_F  n_f \left( -\f{55}{24} +2 \zeta_3 \right) \right. \nn\\
& \left. +C_A n_f \left( -\f{209}{108} +\f{10}{9} \zeta_2 -\f{7}{3} \zeta_3 \right)
- \f{1}{27} n_f^2
\right] \, A^{(1)}\, ,
\end{align}
where $\zeta_n$ is the Riemann $\zeta$ function ($\zeta_2=\pi^2/6$, $\zeta_3=1.202..$).

In order to be able to perform the resummation up to NNLL accuracy,
only the coefficient $B^{(2)}$ is lacking.
There have been in the past several attempts to obtain a
numerical value for this coefficient \cite{Turnock:wy,Dokshitzer:1999sh}.
In this work we will present the analytical result for  $B^{(2)}$.

A direct way of extracting the resummation coefficients consists
in comparing the logarithmic structure of a fixed-order
perturbative calculation of the EEC,
with the expansion of the resummed formula in Eq.~(\ref{CS}).
The expansion up to ${\cal O}(\as^2)$ reads
\begin{align}
\label{expansion}
\f{1}{\sigma_T} \f{d\Sigma^{\rm (res.)}}{ d\cos\chi}=&
\f{1}{4y}\Bigg\{\f{\as}{\pi}
\Big[ -{A}^{(1)} \log y +{B}^{(1)} \Big]
+ \left(\f{\as}{\pi}\right)^2 
\Bigg[ 
 \f{1}{2} \left( {A}^{(1)}\right)^2 \log^3 y
 \nn\\&
+\left(-\f{3}{2} {B}^{(1)}  {A}^{(1)} +
\beta_0 {A}^{(1)} \right) \log^2 y  \nn\\
&+ \left( -{A}^{(2)} -
\beta_0 {B}^{(1)}
 +  \left(  {B}^{(1)} \right)^2
 - {A}^{(1)} {H}^{(1)}  \right) \log y \nn\\
& + {B}^{(2)} 
 +  {B}^{(1)} {H}^{(1)} 
  +2 \zeta_3 ({A}^{(1)})^2 \Bigg] +{\cal O}(\as^3) \Bigg\}\, ,
\end{align}
where we have set $\mu_R=Q$.

An analytic calculation of EEC at NLO (i.e. up to ${\cal O}(\as^2)$)
would allow
the extraction of the coefficients ${A}^{(1)}$, ${B}^{(1)}$, 
${H}^{(1)}$, ${A}^{(2)}$ and ${B}^{(2)}$.
However the full analytic result is not really necessary to this purpose:
it is sufficient to compute its small-$y$ behaviour.

The strategy to obtain the small-$y$ behaviour
is the one applied for a similar calculation in the case of
the transverse-momentum distribution in hadronic
collisions \cite{b2calc}.
The singular behaviour at small $y$ ($q_T$) is dictated by the infrared
(soft and collinear) structure of the relevant QCD matrix elements.
At ${\cal O}(\as)$ this structure has been 
known for a long time \cite{BCM}. In recent years,
the universal functions that control the soft and collinear singularities
of tree-level and one-loop QCD amplitudes at ${\cal O}(\as^2)$
have been computed \cite{tree,1loop}.
By using this knowledge, and
exploiting the simple kinematics of the leading-order subprocess, we were able
to construct {\em improved} factorization formulae that allow the 
control of {\em all} infrared singular regions,
avoiding problems of double counting
 \cite{b2calc}. We have
used these improved formulae to approximate the relevant matrix elements and 
compute the small-$y$ behaviour of EEC in a simpler manner.

Compared with the calculation of Ref.~\cite{b2calc},
in the case of EEC there is an additional
complication.
The definition of EEC in Eq.~(\ref{eecdef}) implies that a sum over
all possible parton pairs has to be performed. Thus
an infrared-finite result can be recovered only after
summing over all the correlations.

For hadron-initiated processes,
the coefficient $B_a^{(2) F}$
is generally dependent on the resummation scheme, and on the
way the resummation formula is actually organized \cite{Catani:2000vq}. 
However, despite these ambiguities, it always has the form \cite{b2calc}
\begin{align}
\label{B2}
B_a^{(2)F}&= -\f{1}{2} \gamma_{a}^{(2)} + \f{1}{2}\beta_0 {\cal A}^{F}_a \, , ~~~a=q,g\, ,
\end{align}
where $\gamma_{a}^{(2)}$ is the coefficient of the $\delta(1-z)$ term in the
two-loop splitting function \cite{2loopns,2loops}.
The second term in Eq.~(\ref{B2}) depends
on the virtual correction to the process $a{\bar a}\to F(q_T,Q^2)$.
Considering the similarity between EEC and
the transverse momentum spectra in hadronic collisions,
it is natural to expect a similar form for the coefficient $B^{(2)}$
for the EEC, modulo possible crossing effects.

More precisely, since the leading-order subprocess which is relevant here
is the production of a $q{\bar q}$ pair, we expect
\begin{equation}
\label{B2eec}
B^{(2)}= -\f{1}{2} \gamma_{q}^{(2)} + C_F \beta_0 X.
\end{equation}
Assuming Eq.~(\ref{B2eec}), a calculation of one of the two colour factors $C_FT_R$ or $C_FC_A$ is sufficient to fix the coefficient $X$ in Eq.~(\ref{B2eec}).
We have computed both the $C_FT_R$ and the $C_FC_A$ contributions
to Eq.~(\ref{expansion}) and found complete agreement with all known results.
Our results are also consistent with Eq.~(\ref{B2eec}) and allow us to fix
\begin{equation}
\label{B2eecf}
B^{(2)}= -\f{1}{2} \gamma_{q}^{(2)} + C_F \beta_0\left(\f{5}{6}\pi^2-2\right)\, ,
\end{equation}
the coefficient $\gamma_q^{(2)}$ being
\begin{equation}
\gamma_{q}^{(2)}= C_F^2\left( \f{3}{8}-\f{\pi^2}{2}+6 \zeta_3 \right) +
 C_F C_A \left( \f{17}{24}+ \f{11 \pi^2}{18}-3 \zeta_3\right) 
- C_F n_f T_R\left( \f{1}{6}+ \f{2 \pi^2}{9}\right) \, .
\end{equation}
In principle, since the contribution to Eq.~(\ref{expansion}) proportional to
the colour factor $C_F^2$ has not been computed,
the result in Eq.~(\ref{B2eecf}) is not fully established.
However, besides the parallel
with transverse momentum distributions in hadronic
collisions, which strongly suggests Eq.~(\ref{B2eec}),
there are two additional arguments that confirm it
\footnote{We also note that our result agrees with the
one guessed
by K.~Clay and S.D.~Ellis \cite{Clay:1995sd}
based on the similarity of EEC to Drell--Yan and, as far as the $C_Fn_F$ part is concerned,
with an independent calculation \cite{dixon}.}
.
The first one relies on the correspondence that should exist between
our coefficient $B^{(2)}$ and the quark coefficient in the non-singlet (NS)
scheme \cite{Catani:2000vq}:
\begin{equation}
\label{B2NS}
B^{(2)}_{q,{\rm NS}}= -\f{1}{2} \gamma_{q}^{(2)} + C_F \beta_0\left(\f{\pi^2}{6}-\f{1}{2}\right)\; ,
\end{equation}
which is expected to directly measure the intensity of collinear radiation from quarks at ${\cal O}(\as^2)$.
We find that the difference between Eqs.~(\ref{B2eecf}) and (\ref{B2NS})
can indeed
be explained as a pure crossing
effect due to an additional factor present in the phase space
in the case of EEC.

Finally, the numerical value of $B^{(2)}$ for $n_F=5$, $B^{(2)}=11.2$, is in good
agreement with the estimate of Ref.~\cite{Dokshitzer:1999sh},
$B^{(2)}\sim 10.7$
\footnote{More precisely, the numerical estimate of Ref.~\cite{Dokshitzer:1999sh} is for the coefficient $G_{21}$, which is related to $B^{(2)}$
 by $B^{(2)}=\f{1}{4} G_{21}+\f{5}{12}C_F n_F-\f{1}{4}\left(\f{67}{6}-\f{\pi^2}{2}\right) C_F C_A-\zeta (3) C_F^2$.}
obtained with the numerical program EVENT2 \cite{Catani:1996vz}.

The resummed component obtained in Eq.~(\ref{CS}) has to be properly matched
to the fixed-order result valid at large $y$.
The matching is performed as follows:
\begin{equation}
\label{match}
\f{1}{\sigma_T}\f{d \Sigma^{\rm (fin.)}}{d\cos\chi}
=\left[\f{1}{\sigma_T}\f{d \Sigma}{d\cos\chi}\right]_{\rm f.o.}
-\left[\f{1}{\sigma_T}\f{d \Sigma^{\rm (res.)}}{d\cos\chi}\right]_{\rm f.o.}
\end{equation}
The first term on the right-hand side of Eq.~(\ref{match})
is the usual perturbative
contribution; it is computed with the numerical program of Ref.~\cite{Catani:1996vz} at a given fixed order (LO or NLO) in $\as$.
The second term is obtained by using the expansion of the resummed component
(see Eq.~(\ref{expansion}))
to the {\em same} fixed order in $\as$.
This procedure guarantees that the right-hand side of Eq.~(\ref{deco})
contains the full information on the perturbative calculation plus resummation
of the logarithmically enhanced contributions to all orders.

As we will show in the next section, our result in Eq.~(\ref{B2eecf}) allows us to perform an excellent
matching between the resummed and perturbative NLO result.

We finally note that the functions $g_i$ are singular as $\lambda\to 1$.
The singular behaviour is related to the presence of the Landau pole
in the QCD running coupling. To properly define the $b$ integration,
a prescription to deal with these singularities has to be introduced.
Here, analogously to what was done in Ref.~\cite{Bozzi:2003jy}, we follow Ref.~\cite{prescription}
and deform the integration contour to the complex $b$-space.

\section{Phenomenological results}

\begin{figure}[htb]
\begin{center}
\begin{tabular}{c}
\epsfxsize=8.5truecm
\epsffile{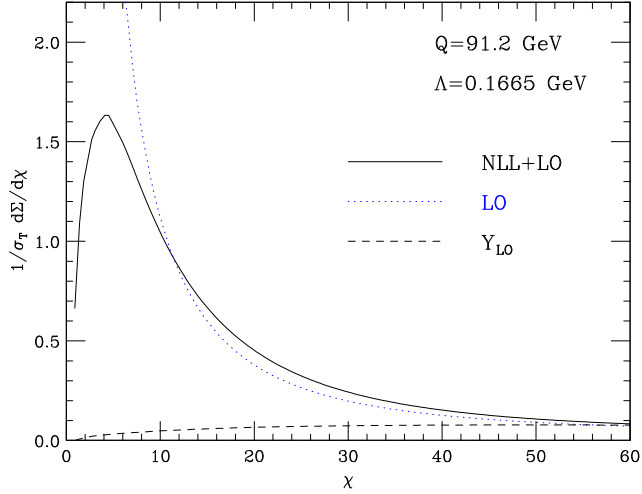}
\end{tabular}
\end{center}
\vspace{-0.7cm}
\caption{\label{fig:nlla}{\em Results to NLL+LO accuracy.}}
\end{figure}
\begin{figure}[htb]
\begin{center}
\begin{tabular}{c}
\epsfxsize=8.5truecm
\epsffile{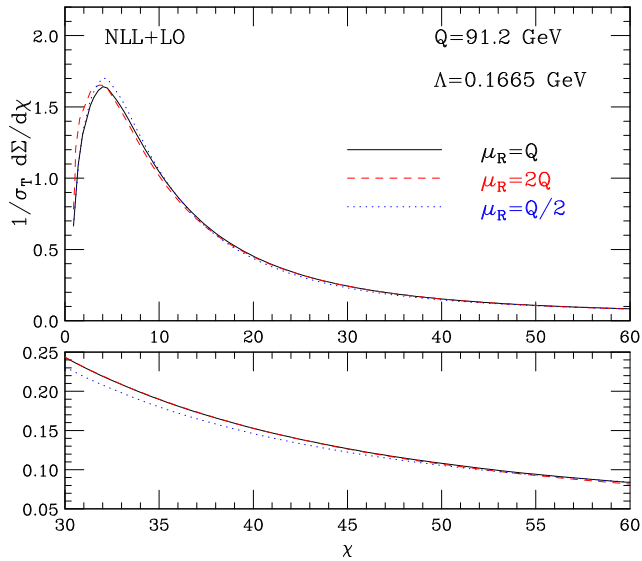}
\end{tabular}
\end{center}
\vspace{-0.7cm}
\caption{\label{fig:nllb}{\em Results to NLL+LO accuracy: scale dependence.}}
\end{figure}

In the following we present quantitative results at NLL+LO and NNLL+NLO
accuracy.
At NLL+LO the resummed component in Eq.~(\ref{CS})
is evaluated by including the functions $g_1$ and $g_2$
in Eq.~(\ref{gfunctions})
and the coefficient $H^{(1)}$ in Eq.~(\ref{H1}).
The finite component in Eq.~(\ref{match}) is instead evaluated at LO and
the one-loop expression for $\as$ is used.
At NNLL+NLO we include also the function $g_3$ in the resummed
component and we evaluate the finite part at NLO, with $\as$ at two-loop level.

The NLL+LO results are shown in Figs.~\ref{fig:nlla}  and \ref{fig:nllb}.
They are obtained by fixing $\Lambda^{n_F=5}_{\rm QCD}=0.1665$ GeV, corresponding to $\as(M_Z)=0.130$.
In Fig.~\ref{fig:nlla} we show the results for $\mu_R=Q$. The dotted line is the LO result, which 
diverges to $+\infty$ as $\chi\to 0$. The solid line is the matched result, and the dashed line gives the matching
term in Eq.~(\ref{match}).
Note that  we plot $1/\sigma_T~d\Sigma/d\chi$, so that the matching term is
actually $Y(\chi)\equiv 1/\sigma_T~d\Sigma^{(\rm fin.)}/d\chi$.
As can be observed, the matching term is well
 behaved  up to very small values of $\chi$ and becomes dominant at 
larger $\chi$, where the fixed-order contribution is expected to control
the matched calculation.

The scale dependence at NLL+LO is studied in Fig.~\ref{fig:nllb}, where the results for the scales $\mu_R=Q/2,Q,2Q$ are shown. The lower plot shows the detail of the region $30^\circ <\chi< 60^\circ$.

\begin{figure}[htb]
\begin{center}
\begin{tabular}{c}
\epsfxsize=8.5truecm
\epsffile{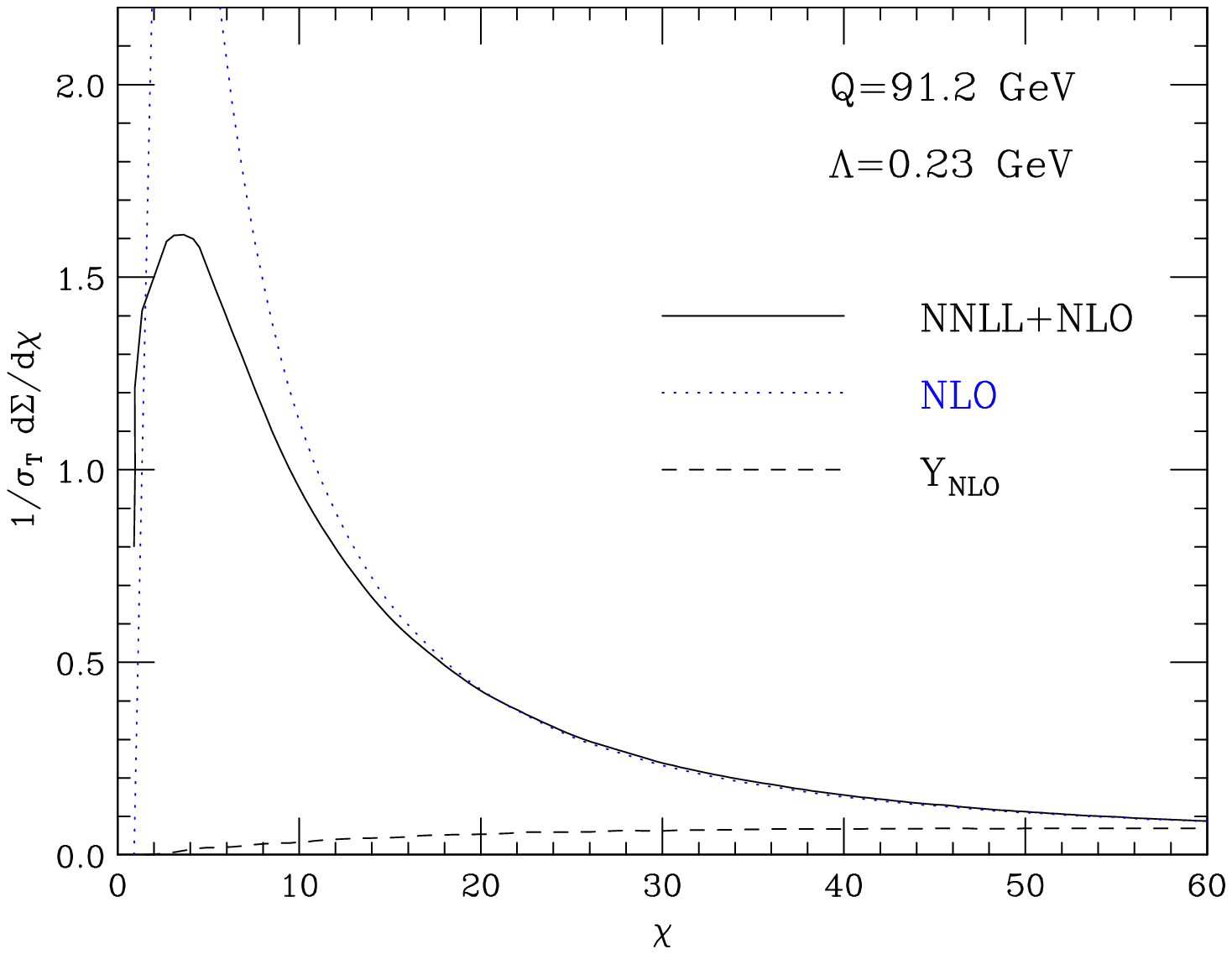}
\end{tabular}
\end{center}
\vspace{-0.7cm}
\caption{\label{fig:nnlla}{\em Results to NNLL+NLO accuracy.}}
\end{figure}
\begin{figure}[htb]
\begin{center}
\begin{tabular}{c}
\epsfxsize=8.5truecm
\epsffile{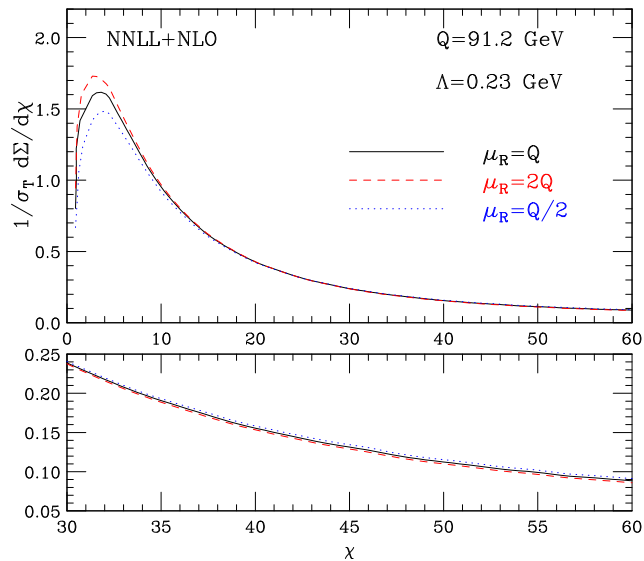}
\end{tabular}
\end{center}
\vspace{-0.7cm}
\caption{\label{fig:nnllb}{\em Results to NNLL+NLO accuracy: scale dependence.}}
\end{figure}

The NNLL+NLO results are shown in Figs.~\ref{fig:nnlla} and \ref{fig:nnllb}.
They are obtained using $\Lambda^{n_F=5}_{\rm QCD}=0.23$ GeV, corresponding to $\as(M_Z)=0.118$.
As before, Fig.~\ref{fig:nnlla} shows the results for $\mu_R=Q$. The dotted line is the NLO contribution, which 
diverges to $-\infty$ as $\chi\to 0$. The solid line is the matched result,
and the dashed line gives the matching
term in Eq.~(\ref{match}).
We see that this matching term displays a very smooth behaviour as $\chi\to 0$, and this is a further
confirmation of the validity of our result in Eq.~(\ref{B2eecf}).
The NNLL+NLO result is plotted in Fig.~\ref{fig:nnllb} for $\mu_R=Q/2,Q,2Q$. As in Fig. \ref{fig:nllb},
the lower panel shows the detail of the region $30^\circ<\chi< 60^\circ$.

We see that scale variations act differently in the low-$\chi$ and medium-$\chi$ regions.
In the region of the peak, lowering (increasing)  $\mu_R$ has the effect of increasing (lowering) $\as(\mu_R)$ and
thus increasing (damping) the Sudakov suppression. This results in the fact that, at NNLL+NLO, the curve at $\mu_R=2Q$
is higher than the one at $\mu_R=Q/2$. This behaviour changes as $\chi$ increases and at $\chi=60^\circ$ the curves are in the
{\em usual} order.
At NLL+LO these two distinct features (Sudakov suppression and
perturbative increase with $\as$)
are less evident, and thus the scale dependence appears smaller than at NNLL+NLO.
At NNLL+NLO the scale dependence is about $\pm 8\%$ at the peak and about $\pm5\%$ at $\chi=60^\circ$, giving an idea of the theoretical uncertainty in the resummed calculation.

We find that the NNLL effect is dominated by the contribution of $B^{(2)}$
in the function $g_3$. By keeping only the term proportional to $B^{(2)}$ in the function $g_3$
in Eq.~(\ref{gfunctions}), the difference with respect to the full NNLL+NLO result is smaller than $1\%$.

\begin{figure}[htb]
\begin{center}
\begin{tabular}{c}
\epsfxsize=9truecm
\epsffile{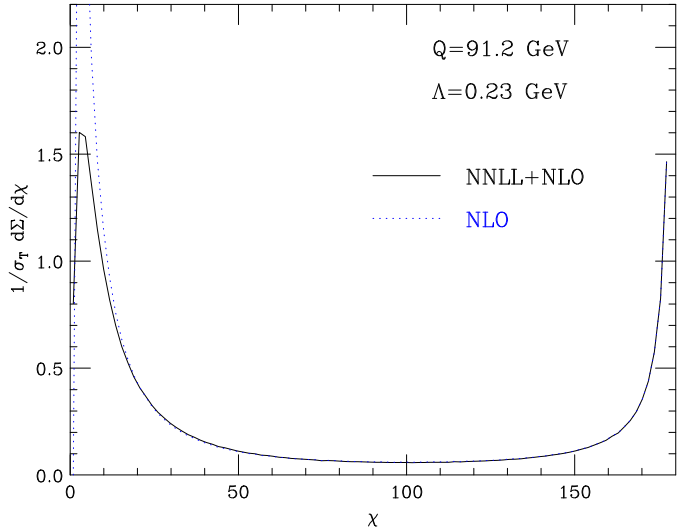}
\end{tabular}
\end{center}
\vspace{-0.7cm}
\caption{\label{fig:osci}{\em NNLL+NLO results in the full $\chi$ range}}
\end{figure}

Figure~\ref{fig:osci} shows the  NNLL+NLO matched
result in the full range of $\chi$.
We see that, contrary to what
happens in other approaches to $b$-space resummation \cite{osci},
there are no oscillations in the medium--high $\chi$ region, where the matched result follows the NLO fixed order calculation.

\begin{figure}[htb]
\begin{center}
\begin{tabular}{c}
\epsfxsize=9truecm
\epsffile{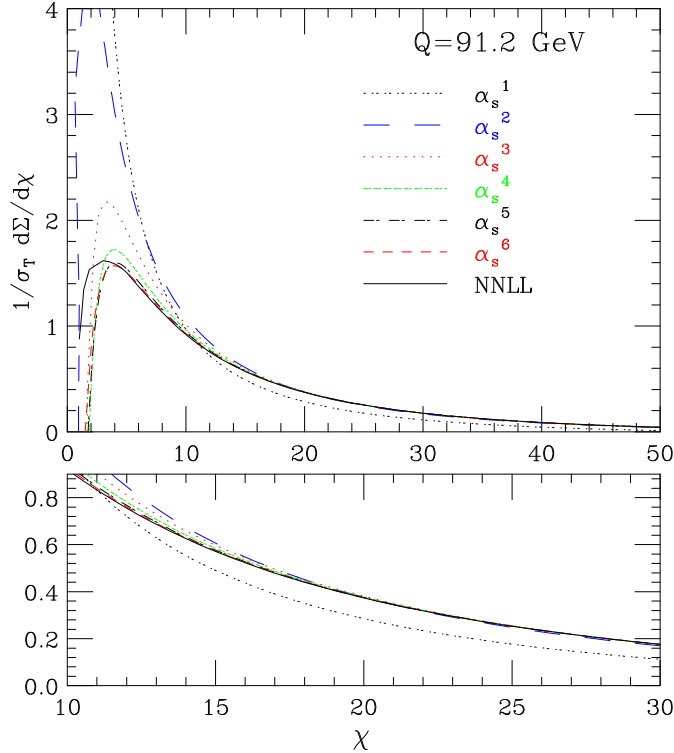}
\end{tabular}
\end{center}
\vspace{-0.7cm}
\caption{\label{fig:exp}{\em Comparison between the purely resummed and the expanded expression (up to ${\cal O}(\as^6)$) for the EEC  }}
\end{figure}

Before moving to the comparison with the experimental data,
we want to study the convergence of the resummed expression, as a check of
the validity of the prescription introduced in \cite{prescription} and used to deal with the Landau pole.
In Fig.~\ref{fig:exp} we compare the purely resummed result
in Eq.~(\ref{CS}) at NNLL accuracy (that is
with the functions $g_1$, $g_2$, $g_3$
and the coefficient $H^{(1)}$ included)
with its expansion up to
${\cal O}(\as^6)$.
As can be observed,
the expansion converges very rapidly to the resummed result in the region of medium  $\chi$, confirming the validity of the prescription where the
fixed-order result dominates (see the lower plot for a detailed comparison).
Nonetheless,
even though the higher the expansion the better the agreement
with the resummed result at smaller values of $\chi$,
for $\chi\ltap 10^\circ$
the fixed-order result is bound to fail, no matter
how many orders in perturbation theory are included,
thus requiring the resummation to all orders.

We can now perform a comparison of the most accurate theoretical NNLL+NLO results with the  precise OPAL \cite{Acton:1993zh} and SLD \cite{Abe:1994mf} data.
As the extraction of the strong coupling constant $\as$ is one of the main
motivations for the measurement of event-shape observables,
we perform a fit of the experimental data on EEC leaving $\Lambda_{\rm QCD}$
as a free parameter.
We do not attempt to produce the most accurate extraction
of $\as(M_Z)$, since we cannot properly take into account
 correlations between the data points and therefore just
add systematic and statistical errors in quadrature.
 
For the moment we neglect hadronization effects,
but we will come back to this point below.
The reader should keep in mind that the
results obtained without including those effects should be considered
with care.

In a first sample we include data in the range $15^\circ< \chi < 120^\circ$
and fix the renormalization scale to $\mu_R=Q=M_Z$.
The upper limit is chosen so as to cut the large angle region
where another resummation would be required.

The quality of the fit is poor, as can be determined by the
value of $\chitwodof=5.17$,
with a rather large value for the coupling constant $\as(M_Z)=0.133\pm0.002$,
in agreement with similar results found by OPAL \cite{Acton:1993zh}.
The uncertainty is dominated by missing higher-order contributions,
estimated by repeating the fit with $\mu_R=1/2\,(2) M_Z$.
Better fit results are actually found when the renormalization scale is also varied, with a reduction of a factor of 2 in $\chi^2$ when $\mu_R\simeq M_Z/2$ and a slightly lower preferred value of the coupling constant.
\begin{figure}[htb]
\begin{center}
\begin{tabular}{c}
\epsfxsize=9truecm
\epsffile{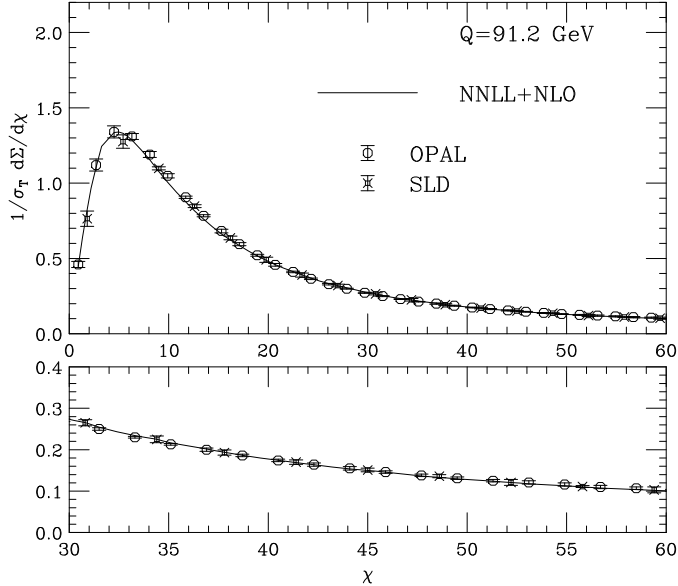}
\end{tabular}
\end{center}
\vspace{-0.7cm}
\caption{\label{fig:ptfit}{\em NNLL+NLO: purely PT fit in the low-$\chi$ region}}
\end{figure}
In a second attempt, we include data in the range $0^\circ< \chi< 63^\circ$,
to isolate the region where the effects of the resummation are considerably more significant.
Even with the scale fixed to  $\mu_R=M_Z$, a very reasonable value
of $\chitwodof= 1.67$ is found, corresponding to
$\as(M_Z)=0.131\pm 0.002$. The error is again dominated by
scale variations $\mu_R=1/2 M_Z,2M_Z$.
The nice comparison between the resummed
calculation and the data in the low-angle
region is shown in Fig.~\ref{fig:ptfit}. It is clear that our NNLL+NLO
result for
the EEC can reproduce very well the data up to the lowest measured angle.

Since the low-$\chi$ region is particularly sensitive to non-perturbative (NP)
effects, whereas in the large angle region we may expect non negligible higher-order (NNLO) contributions, we have repeated the fit in the range
$15^\circ< \chi< 63^\circ$. The value of  $\chitwodof$
goes down to 0.66 but the
result for $\as(M_Z)$ does not change significantly.

We have also investigated the possible effect of the unknown second-order
coefficient $H^{(2)}$ (see Eq.~(\ref{cexp})),
by letting it vary together with $\as(M_Z)$ in
the range $0^\circ < \chi< 63^\circ$. The results show that data still prefer
a high $\as(M_Z)$ and a relatively small $H^{(2)}$.

Up to now we have considered only the perturbative contribution
in the theoretical calculation.
However, NP contributions are expected to be relevant, particularly
for small angles \cite{Basham,CollinsSoper,Fiore:1992sa}.
Thus, following Ref.~\cite{Dokshitzer:1999sh},
we include NP effects by supplementing
the Sudakov form factor in Eq.~(\ref{sudakov}) with a correction of the form
\begin{equation}
\label{NPform}
S_{\rm NP}=e^{-\f{1}{2}a_1 b^2}(1-2a_2 b)\, .
\end{equation}
We have performed a three-parameter preliminary fit to the data still in the
range $0^\circ<\chi< 63^\circ$. We find that the data prefer very small values of
the NP coefficient $a_2$, $|a_2|\ltap 0.002$.
We have thus set $a_2=0$ to perform
the fit.
We obtain for  $\as(M_Z)$ and $a_1$
the following result:
\begin{equation}
\as(M_Z)=0.130^{+0.002}_{-0.004}~~~~a_1=1.5^{+3.2}_{-0.5}~{\rm GeV}^2
\end{equation}
with $\chitwodof=0.99$. 
The error is dominated by scale uncertainty, which is estimated as above
by repeating the fit with $\mu_R=1/2(2) M_Z$.

We see that the quality of the fit improves, but the value
of $\as(M_Z)$ still remains high with respect to the world average
$\as(M_Z)=0.1182 \pm 0.0027$ \cite{bethke}.
Moreover, the three-parameter fit suggests that in our approach the NP
coefficient $a_1$ is more important than $a_2$. We conclude that
the parametrization in Eq.~(\ref{NPform})
is not able to fully take into account
the hadronization effects,
particularly at medium and large values of $\chi$.
The extracted ``effective'' coupling $\as$ thus 
absorbs part
of the hadronization effects.
We stress that this result is not an artefact
of the resummation procedure, since a similar effect is observed
when the large-angle data are compared with
the fixed order NLO result.

A different method
to include NP (hadronization) corrections,
extensively applied in the QCD analysis to event-shape variables at LEP,
is to use a parton shower
Monte Carlo. This method is certainly model-dependent, since
different approaches
have been developed to describe the hadronization process,
but has the advantage that
the free parameters
are tuned through fits to large sets of different data.

In Refs.~\cite{Acton:1993zh,Abe:1994mf} the data for EEC have been corrected
at parton level before performing the QCD analysis.
We have used the parton-level data of Ref.~\cite{Acton:1993zh} to repeat
our fit.
The hadron--parton correction factors are large, from about $1.5$ in the very small $\chi$ region
to  $\sim0.9$ at large $\chi$.
The quality of the fit in terms of $\chitwodof$ is generally worse than
before, but
errors related to
the hadronization correction have not been included.
The uncertainty from hadronization is usually estimated by
trying different alternatives for the hadronization correction, and using
the spread in the results as an estimate of the ensuing error.
In Ref.~\cite{Acton:1993zh} the hadronization
uncertainty on $\as(M_Z)$ is estimated to be about $\pm 4\%$.

\begin{figure}[htb]
\begin{center}
\begin{tabular}{c}
\epsfxsize=9truecm
\epsffile{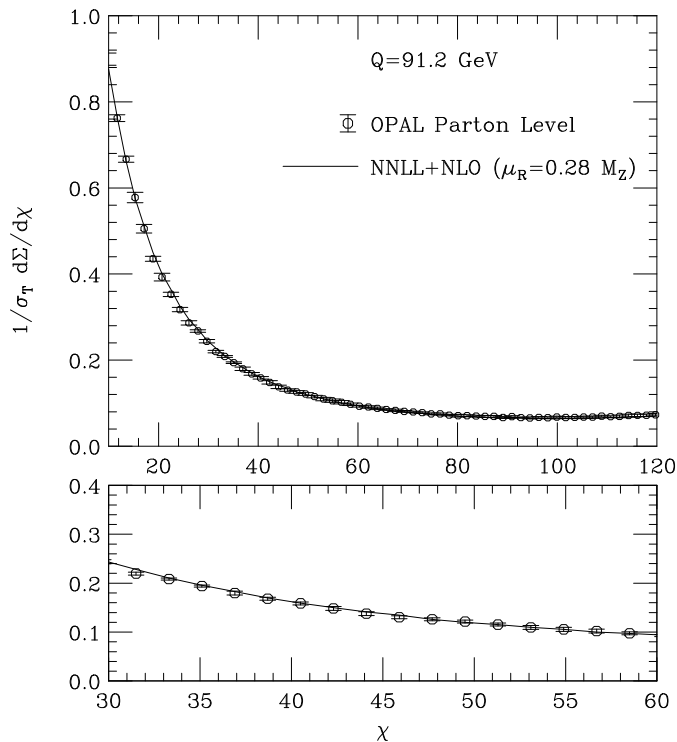}
\end{tabular}
\end{center}
\vspace{-0.7cm}
\caption{\label{fig:parton}{\em Comparison of NNLL+NLO fit to parton level OPAL data with $\as(M_Z)=0.1175$ }}
\end{figure}

In the kinematical range $15^\circ <\chi< 63^\circ$ we find
$\chitwodof=3.78$ and $\as=0.119\pm 0.001$, with a very similar result
($\chitwodof=5.02$ and $\as=0.120\pm 0.001$) when the fit range is extended to $15^\circ <\chi<120^\circ$.
Finally, when the fit is repeated by
allowing the variation of the renormalization scale, 
excellent results are obtained,
the best fit corresponding to $\as=0.1175$,
 $\chitwodof=1.36$ and $\mu_R=0.28\, M_Z$ (see Fig. \ref{fig:parton}).
Thus, when hadronization effects are taken into account using
a Monte Carlo, as done in Refs.~\cite{Acton:1993zh,Abe:1994mf},
the results we obtain for $\as(M_Z)$ are in complete agreement with the world
average.

We note that
the QCD analysis performed by OPAL  on the same parton-level data
gave, instead, $\as(M_Z)=0.132^{+0.008}_{-0.007}$ \cite{Acton:1993zh},
considerably higher than our result.
The analysis
of Ref.~\cite{Acton:1993zh} (as well as the one of Ref.~\cite{Abe:1994mf})
used the NLL resummed calculation of
Ref.~\cite{Turnock:wy}.
In order to understand the origin of the discrepancy,
we have compared our results with those of Ref.~\cite{Turnock:wy}.
We find that
the differences are not negligible, especially in the region
$\chi\ltap 40^\circ$,
and may explain the above discrepancy.
The approach of Ref.~\cite{Turnock:wy} was based
on an approximated analytic
evaluation
of the $b$-space integral of Eq.~(\ref{CS}),
and suffers from an unphysical singularity at very small $\chi$.
The effect of this singularity may propagate also within the range of the
fit, thus spoiling the resummed prediction.
The other source of difference
is that the coefficient $B^{(2)}$,
or equivalently $G_{21}$,
was evaluated numerically, thus leading to a larger uncertainty in
the matching procedure.

\section{Conclusions}

In this paper we have
considered the observable known as energy--energy correlation in
$e^+e^-$ collisions.
We have provided the complete structure
of the ${\cal O}(\as^2)$ logarithmically enhanced contributions up to
NNLL accuracy,
by giving the expression of the unknown second-order coefficient $B^{(2)}$,
needed to reach full NNLL precision.

We have presented perturbative predictions both at NLL+LO and NNLL+NLO
accuracy, by showing that the knowledge of the
calculated coefficient $B^{(2)}$ allows us to perform an excellent matching
of the resummed and fixed-order calculations.
We have studied the  impact of the results in a detailed comparison to precise
LEP and SLC data. A good description of the data is obtained
but the extracted value of $\as(M_Z)$
turns out to be high when hadronization effects are neglected
or parametrized using the form
in Eq.~(\ref{NPform}).
By contrast,
using OPAL data corrected at parton level, that were obtained
by estimating hadronization corrections
using a Monte Carlo parton shower, the values
of $\as(M_Z)$ we find are in good agreement with the current world average.

\noindent {\bf  Acknowledgements}

\noindent
We would like to thank Stefano Catani for many helpful
discussions and suggestions. We also thank Mrinal Dasgupta, Lance Dixon,
Luca Trentadue and Bryan Webber for useful discussions and comments.
We are grateful to David Ward and the OPAL collaboration
for providing us the parton-level data used in the analysis.

\end{document}